\documentclass{SciPost}
\hypersetup{
    colorlinks,
    linkcolor={red!50!black},
    citecolor={blue!50!black},
    urlcolor={blue!80!black}
}

\usepackage[bitstream-charter]{mathdesign}
\usepackage{orcidlink}
\DeclareSymbolFont{usualmathcal}{OMS}{cmsy}{m}{n}
\DeclareSymbolFontAlphabet{\mathcal}{usualmathcal}

\fancypagestyle{SPstyle}{
\fancyhf{}

\fancyfoot[C]{\textbf{\thepage}}
}

\usepackage{url}

\usepackage[scaled=0.9]{sourcecodepro}
\usepackage[svgnames]{xcolor}
\definecolor{bgclr}{RGB}{240, 240, 240}
\definecolor{commentclr}{RGB}{61, 122, 122}
\definecolor{keywordclr}{RGB}{51, 51, 204}
\definecolor{moduleclr}{RGB}{82, 0, 82}
\definecolor{typeclr}{RGB}{205,102,0}
\definecolor{stringclr}{RGB}{153, 51, 51}
\usepackage{listings}
\lstdefinestyle{GlobalStyle}{
    frame=none,
    backgroundcolor=\color{bgclr},
    basicstyle={\fontsize{9.5}{11.5} \selectfont\ttfamily},
    keywordstyle=[1]\color{keywordclr},
    keywordstyle=[2]\color{keywordclr},
    keywordstyle=[3]\color{typeclr},
    keywordstyle=[4]\color{moduleclr},
    commentstyle=\color{commentclr},
    stringstyle=\color{stringclr},
    breaklines=true,
    tabsize=4,
    numbers=left,
    numberstyle=\tiny,
    stepnumber=1,
    numbersep=6pt,
    firstnumber=1,
    captionpos=b
}
\lstdefinelanguage{falconcode}{
    morekeywords=[1]{},
    morekeywords=[2]{terminal, if, else, elif, import, autotuner, struct, state, start, routine},
    morekeywords=[3]{Error, string, int, Array, DeviceConnection, DeviceStates, Config},
    sensitive=true,
    morecomment=[l]{//},
    morestring=[b]",
    literate={->}{{{\color{keywordclr}{->}}}}{1}
}
\lstdefinelanguage{json}{
    morestring=[b]",
}
\usepackage{xparse}
\NewDocumentCommand{\codelines}{m g}{%
  \IfNoValueTF{#2}{line~#1}{lines~#1--#2}%
}

\usepackage{hyperref}

\definecolor{trailer}{gray}{0.9}
\def\formtmp#1#2{{\vskip10pt\noindent\fboxsep=0pt\colorbox{#1}{\vbox{\vskip3pt\hbox to \textwidth{\hskip3pt\vbox{\raggedright\noindent\textbf{#2\vphantom{Qy}}}\hfill}\vspace*{3pt}}}\par\vskip2pt%
\noindent\kern0pt}}
\newenvironment{trailer}[1]{\ignorespaces\def\stmtopen##1{##1}%
\formtmp{trailer}{#1}}{\par\noindent\textcolor{trailer}{\rule{\columnwidth}{1pt}}\vskip2pt\par\addvspace{0.5\baselineskip}}%

\usepackage{dsfont}
\usepackage{siunitx}

\usepackage[T1]{fontenc}
\usepackage[scaled=1.0]{beramono}  

\begin{document}
\pagestyle{SPstyle}
\begin{center}{\Large \color{scipostdeepblue}{
\texttt{FAlCon}: \textbf{A unified framework for algorithmic control of quantum dot devices\\
}}}\end{center}
\begin{center}\textbf{
Tyler J. Kovach\orcidlink{0009-0007-0807-7300}\textsuperscript{1$\dagger\star$},
Daniel Schug\orcidlink{0009-0001-3758-501X}\textsuperscript{2,3,4$\star$},
Zach D. Merino\textsuperscript{4},
Mark Friesen\orcidlink{0000-0003-2878-2844}\textsuperscript{1},\\
Mark A. Eriksson\orcidlink{0000-0002-3130-9735}\textsuperscript{1},
Justyna P. Zwolak\orcidlink{0000-0002-2286-3208}\textsuperscript{3,4,5$\ddagger$}
}\end{center}

\begin{center}
{\bf 1} Department of Physics, University of Wisconsin-Madison, Madison, WI 53706, USA
\\
{\bf 2} Department of Computer Science, University of Maryland, College Park, MD 20742, USA
\\
{\bf 3} Joint Center for Quantum Information and Computer Science,
University of Maryland, College Park, MD 20742, USA
\\
{\bf 4} National Institute of Standards and Technology, Gaithersburg, MD 20899, USA
\\
{\bf 5} Department of Physics, University of Maryland, College Park, MD 20742, USA
\\[\baselineskip]
$\dagger$ \href{mailto:tkovach@wisc.edu}{\small tkovach@wisc.edu}\,\quad
$\ddagger$ \href{mailto:jpzwolak@nist.gov}{\small jpzwolak@nist.gov}\,\quad
$\star$ These authors contributed equally.
\end{center}
\section*{\color{scipostdeepblue}{Abstract}}
\textbf{\boldmath{%
As spin-based quantum systems scale, their setup and control complexity increase sharply.
In semiconductor quantum dot (QD) experiments, device-to-device variability, heterogeneous control-electronics stacks, and differing operational modalities make it difficult to reuse characterization, calibration, and control logic across laboratories.
We present \texttt{\normalfont{FAlCon}}, an open-source software ecosystem for portable, automated characterization and tuning measurement workflows.
\texttt{\normalfont{FAlCon}} provides (i) a lightweight domain-specific language for expressing state-based tuning logic in a hardware-agnostic form; (ii) specialized transmittable libraries of physics-informed QD data structures (``tuning vernacula''); and (iii) extensible libraries of shared measurement protocols enabling an interoperable lab-agnostic measurement stack.
By separating algorithm intent from instrument realization, while preserving traceability and supporting typed scripting, \texttt{\normalfont{FAlCon}} enables researchers and engineers to exchange, adapt, and deploy characterization and autotuning routines across heterogeneous QD setups.
The framework supports all users, ranging from end users running prebuilt algorithms with custom initial conditions to developers extending instrumentation support and contributing new tuning strategies.
Although the present release targets QD experiments, other qubit modalities and scientific experiments could reuse \texttt{\normalfont{FAlCon}}'s modular abstractions by providing new tuning data types and instrument control templates.
}}

\vspace{\baselineskip}
\vspace{5pt} 
\noindent\rule{\textwidth}{1pt}
\vspace{-20pt}
\tableofcontents
\noindent\rule{\textwidth}{1pt}
\vspace{5pt}

\section{Introduction}
\label{sec:intro}
Semiconductor quantum dots (QDs) are a leading platform for spin-based quantum information processing, offering dense quantum-bit (qubit) packing and compatibility with state-of-the-art semiconductor manufacturing~\cite{Loss98-QCD, Vandersypen17-ISQ, Vandersypen19-QCS, Burkard21-SSQ, George24-QAF, Steinacker25-ISS}.
As experiments scale from few-dot devices to extended one- and two-dimensional (1D and 2D) arrays containing tens or hundreds of quantum dots, the effort required to characterize devices, calibrate operating conditions, and maintain stability becomes a dominant bottleneck to the formation and operation of physical qubits.
This operational workload spans device bootstrap, charge-state initialization, tuning of tunnel barriers and interdot couplings, charge-sensor calibration and readout configuration, qubit initialization, quantum gate characterization, and continual drift compensation across multiple voltage subspaces and timescales.
In many laboratories, tuning remains a largely manual process that requires expert intervention, becoming increasingly impractical as device sizes grow.

One of the central challenges in QD device operation is strong device-to-device variability~\cite{Vandersypen17-ISQ, Burkard21-SSQ}.
Variations in heterostructures, material disorder, and device architectures mean that no two devices exhibit identical responses or admit identical tuning trajectories. 
As a result, tuning procedures that succeed on one sample often require substantial adaptation when transferred to a new device, even within the same material system. 
This challenge is compounded by the absence of standardized control-electronics stacks, the diversity of measurement styles, and the differing internal wiring of dilution refrigerators, which leads to widely varying software interfaces and measurement pipelines.
Consequently, high-level characterization and tuning logic that could in principle be shared is frequently rewritten against local instrument application programming interface (API) and lab-specific scripting conventions, limiting reuse across groups and slowing community-level iteration.

These difficulties have motivated substantial progress on automation of QD characterization and tuning~\cite{Zwolak21-AAQ}. 
Algorithmic and rule-based approaches have demonstrated reliable navigation to target charge regimes and efficient adjustment of operating parameters~\cite{Baart16-CAT, vanDiepen18-ATC, Lapointe-Major19-ATQ}.
Machine-learning methods have further enabled robust state recognition, feature extraction, and in situ autotuning under realistic measurement noise and device variability~\cite{Kalantre17-MLD, Ziegler22-TRA, Zwolak20-AQD, Durrer19-ATQ}.
However, many successful demonstrations remain tightly coupled to specific experimental control stacks, data representations, and instrumentation interfaces, making it difficult to exchange tuning routines across laboratories without substantial reimplementation.

In this work, we introduce \texttt{FAlCon} (a Framework for ALgorithmic CONtrol), an open-source software ecosystem designed to address this portability gap in the QD community.
\texttt{FAlCon} is a software stack that (1) defines a high-level domain-specific language (DSL) for writing autotuning algorithms, (2) defines an extensive library of physics-informed data structures, and (3) connects those algorithms to a lab-agnostic instrument-facing interface.
\texttt{FAlCon} enables characterization, calibration, and tuning routines to be expressed in a hardware-agnostic form and deployed on heterogeneous laboratory setups.

From a software engineering perspective, \texttt{FAlCon} is implemented as a suite of software rather than a monolithic application.
At the foundation, there is a defined core library of serializable data structures that express physics-based QD tuning concepts.
Above this library, language bindings enable tools and services to be implemented in multiple programming environments without re-implementing core semantics.
Across the libraries and applications, there is a well-defined inter-language communication protocol for sending and receiving any core library data structures. 
This enables them to be interchanged with alternative compliant implementations, which together compose a full compliant software suite.

Although the present implementation targets QD experiments, \texttt{FAlCon}'s modular abstractions are designed to generalize to other scientific use cases by providing new data types in the specified domain and instrument templates.
The framework is intended to support a wide range of users, from experimentalists running prebuilt routines with custom initial conditions to developers contributing new tuning strategies and extending instrumentation support. 
Such capabilities will become increasingly important as QD processors scale to larger arrays requiring extensive calibration and continual automated retuning.
In the remainder of this article, we describe the architecture of \texttt{FAlCon} and its use in the context of QD autotuning.
We also discuss how it can be extended to additional data types and different instrument backends.

\texttt{FAlCon} is available for download from GitHub, with the source code released under the Mozilla Public License (MPL) 2.0 and the BSD 3-Clause BSDv3 License.
The current release is organized as seven public repositories within the \texttt{FAlCon} GitHub organization~\cite{FAlCon}.
\begin{trailer}{\texttt{falcon-autotuning}} 
\texttt{falcon-autotuning} defines the organization-level control layer of \texttt{FAlCon}.
\end{trailer}
\begin{trailer}{\texttt{falcon-lib}} 
\texttt{falcon-lib} contains the DSL for encoding autonomous control and optimization algorithms as state machines (in C++), the associated runtime tooling for parsing, interpreting, and deploying these routines (in C++), and initial libraries for the language (in the falcon DSL).
\end{trailer}
\begin{trailer}{\texttt{falcon-core}}
\texttt{falcon-core} provides the performance-oriented foundation (primarily in C++ with a C API) for canonical data types and portable serialization used across the ecosystem.
\end{trailer}
\begin{trailer}{\texttt{falcon-core-libs}}
\texttt{falcon-core-libs} exposes \texttt{falcon-core} capabilities through multi-language bindings (currently including Python, Go, OCaml, and Lua) to support tooling and integration in different programming environments.
\end{trailer}
\begin{trailer}{\texttt{falcon-instrument-hub}}
\texttt{falcon-instrument-hub} provides an abstracted interface layer (primarily in Go) for falcon measurement request parsing, instrument execution command requests, and measurement data transfer.
\end{trailer}
\begin{trailer}{\texttt{instrument-script-server}}
\texttt{instrument-script-server} implements the lab-facing execution and orchestration service (primarily in C and C++, with Lua-based scripting support), providing process-isolated instrument workers, plugin-based drivers, synchronization across instruments, and structured result capture.
\end{trailer}
\begin{trailer}{\texttt{falcon-measurement-lib}}
\texttt{falcon-measurement-lib} defines a schema-driven measurement interface layer (JSON Schema with generators and supporting code in Go and Lua/Teal), standardizing script contexts and reusable types and enabling code generation to support safer script development and validation.
\end{trailer}
\noindent
Together, these repositories separate algorithmic intent from instrument implementation while providing a practical path to deploy portable state-based routines across heterogeneous QD laboratory setups.

\section{Motivation}
\label{sec:motivation}
From an experimentalist's perspective, a recurring challenge in autotuning is not the design of high-level logic, but the distribution and reuse of the resulting code.
In the QD community, tuning software is often inseparable from the experimental stack in which it was developed.
In practice, an ``autotuning algorithm'' often implicitly depends on the local instrument interfaces, measurement orchestration, data formats, calibration conventions, and assumptions about acquisition timing and hardware capabilities.
As a result, a tuning routine that is reliable in one setup is rarely reusable as a drop-in component in another, even when the physical tuning goals are similar.

\begin{figure}[!t]
    \centering
    \includegraphics[width=0.8\linewidth]{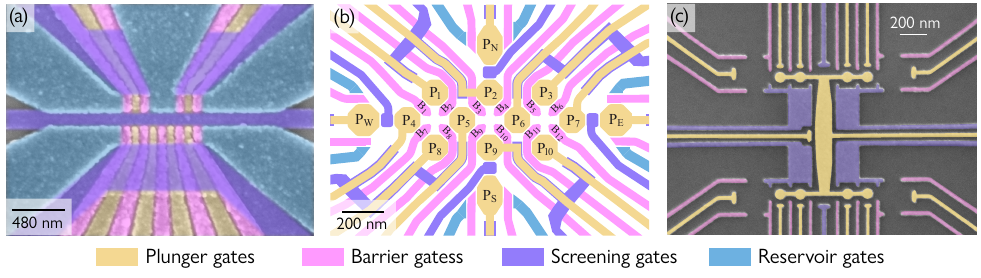}
    \caption{Examples of three gate-defined QD device architectures illustrating the diversity of layouts and the common functional roles of gates (reservoir, screening, plunger, and barrier).
    In principle, each of these systems can be tuned to regimes hosting QDs suitable for spin-qubit operation.
    (a) A SiGe QD device~\cite{Kovach24-BATIS}, hosting up to four qubit QDs in the bottom channel and two charge sensors in the top channel.
    (b) A Ge QD device~\cite{Stehouwer24-ESG}, hosting up to ten qubit QDs in the center region between the screening gates and four charge sensors on the outside of the array.
    (c) A GaAs QD device~\cite{Zubchenko24-ABQ},  hosting up to eight possible QDs organized into four double-QDs, and four charge sensors adjacent to each pair.
    }
    \label{fig:devices}
\end{figure}

Figure~\ref{fig:devices} illustrates the diversity of QD device designs that coexist in current research.
The devices differ substantially in heterostructure, gate layout, and operating modality, including accumulation- and depletion-style operation.
Figure~\ref{fig:devices}(a) shows an accumulation-style device in which gate electrodes are naturally grouped by function (reservoir, screening, plunger, and barrier).
Figure~\ref{fig:devices}(b) shows a higher-density gate architecture on a different accumulation-mode device, with many independently addressable plunger electrodes and multiple reservoirs and screening gates.
Figure~\ref{fig:devices}(c) shows a depletion-mode device style with a distinct lithographic layout and wiring geometry.
Despite these differences, many tuning workflows share common motifs, including establishing stable charge confinement, configuring charge sensing, and iteratively adjusting gate voltages to reach target operating points.

In practice, however, sharing algorithms is difficult because the portability boundary extends beyond the device itself to encompass surrounding instrumentation, control software, and data infrastructure.
A routine implicitly assumes specific instrument interfaces and measurement primitives, a particular orchestration model (e.g., interactive control versus staged sequences), and conventions for storing and interpreting results.
Porting code, therefore, often reduces to rebuilding the surrounding software environment or forking the implementation and adapting it to a new stack.
Representative examples of this issue are the two bootstrapping algorithms introduced in Ref.~\cite{Kovach24-BATIS} and Ref.~\cite{Zubchenko24-ABQ}.
While both approaches aim to bring devices into useful regimes, their implementations are not directly interchangeable because they are tightly coupled to distinct experimental stacks and data representations.
The consequence is duplicated engineering effort and reduced opportunity for systematic comparison of algorithms across platforms.

This portability gap impacts not only algorithm developers but also experimental end users.
End users often wish to study how a controlled physical change (e.g., heterostructure stack, gate geometry, strain, or operating modality) affects device behavior while holding the tuning procedure fixed.
When tuning code cannot be reused, these users must invest effort in rebuilding or reinventing the tooling required to reach the desired operating regime, and they typically implement only what is necessary for a single setup.
This reinforces fragmentation, leaving tuning software local to individual laboratories, minimally generalized, and difficult to benchmark or extend.

\texttt{FAlCon} is designed to address this problem by making autotuning routines portable across heterogeneous QD experimental stacks.
It allows algorithm developers to publish routines in a hardware-agnostic form and end users to execute them across varying laboratory setups by supplying instrument-specific templates and configurations.
By separating algorithm intent from instrument realization, \texttt{FAlCon} aims to reduce duplicated engineering work, improve reproducibility of tuning workflows, and enable more direct comparison of autotuning strategies across devices and laboratories.

\begin{figure}[!t]
    \centering
    \includegraphics[width=\linewidth]{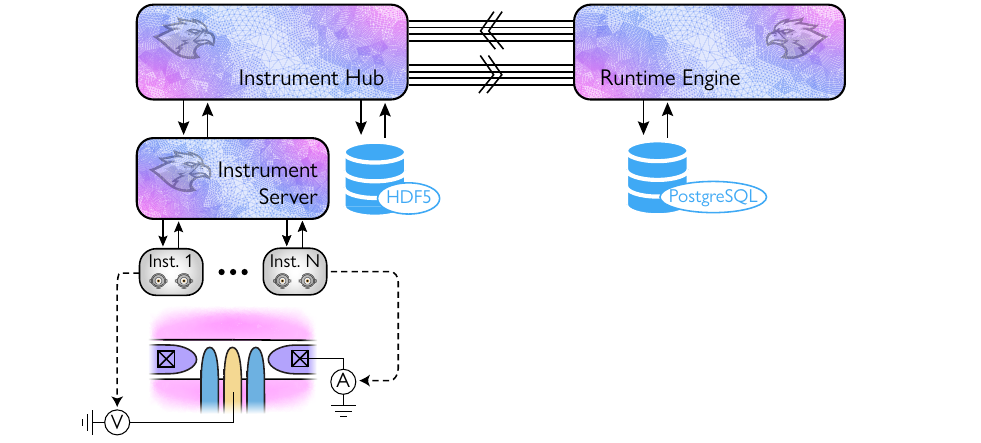}
    \caption{
    High-level architecture of the \texttt{FAlCon} control and measurement stack.
    The Runtime Engine (right) executes \texttt{FAlCon} routines, maintains tuning state and global variables, and persists structured metadata in a PostgreSQL database.
    Measurement requests and results are exchanged with the laboratory-side measurement stack via message passing (e.g., over NATS).
    The Instrument Hub (left) receives measurement requests from the Runtime Engine and translates them into concrete actions executed by the Instrument Server.
    The Instrument Server manages instrument lifetimes and distributes requests across one or more connected instruments (Inst.~1, $\dots$, Inst.~N), which directly interface with the experimental hardware.
    The Instrument Hub aggregates returned results and stores measured datasets as portable HDF5 artifacts~\cite{HDF5}.
    The bottom schematic illustrates a representative device-level wiring context for QD measurements, with a voltage source applied to a gate (\si{\volt}) and a current readout at an Ohmic contact (\si{\ampere}).
    }
    \label{fig:FAlCon-structure}
\end{figure}

\section{\texttt{FAlCon} Framework}
\label{sec:FAlCon-framework}
\texttt{FAlCon} is an open-source ecosystem composed of several related repositories that collectively support portable characterization, calibration, and autotuning workflows for QD experiments.
The ecosystem is intentionally modular, allowing individual components to evolve as new devices, instrument stacks, and user requirements emerge.

At a high level, \texttt{FAlCon} naturally subdivides into a \textit{measurement execution} server and an \textit{algorithmic decision-making} runtime\footnote{Here and thereafter by runtime we mean the period when a program is executing.}.
The measurement execution server handles instrument control, sequencing, and data acquisition at the laboratory level.
The algorithmic decision-making runtime maintains algorithm state, interprets results, and selects the next actions in a tuning routine.
These two constituents communicate bidirectionally so that algorithm decisions can be conditioned on the latest measurements and so that measurement execution can report status, failures, and metadata back to the algorithm layer.

This division is motivated by practical constraints of low-noise cryogenic experiments and by the increasing computational cost of data-driven analysis.
Modern autotuning workflows frequently incorporate machine-learning models to interpret measurements and guide subsequent actions~\cite{Zwolak20-AQD, Zwolak21-AAQ}.
Model inference and associated preprocessing can require substantial CPU and memory resources, and in some cases, benefit from GPU acceleration.
Consolidating heavy computation on the measurement computer is often undesirable in practice, both to minimize local electromagnetic interference and to keep the acquisition stack lightweight and reliable.
\texttt{FAlCon} supports running the algorithm runtime on separate hardware from the measurement service, potentially in a different physical location, while maintaining low-latency coordination through message passing.
In our reference implementation, this coordination is implemented using the NATS messaging system~\cite{NATS}.
Figure~\ref{fig:FAlCon-structure} summarizes this decomposition and the primary communication paths. 

The \texttt{FAlCon} ecosystem is distributed as a GitHub organization that serves as the public entry point and as the current set of open-source repositories~\cite{FAlCon}.
The organization's landing page provides a high-level description of the project goals, development status, community expectations, and links to the main repositories. 
In the remainder of this paper, we use ``\texttt{FAlCon}'' to refer to the overall measurement-and-control ecosystem rather than any single repository.
In the following subsections, we describe the main repositories in the current release and their roles within the overall framework.

\subsection{\texttt{falcon-lib}}
\label{ssec:falcon-org}
At a conceptual level, \texttt{falcon-lib} separates a \textit{control-plane layer}, which represents and executes tuning logic, from a \textit{measurement-plane layer}, which interfaces with laboratory instruments and acquires data.
The control plane is centered on a user-facing DSL for expressing tuning routines as hierarchical, state-based programs.
A corresponding runtime engine parses and executes DSL programs while exposing controlled interfaces to external services and analysis libraries.
In the reference implementation, bidirectional communication between the control and measurement services is implemented via message passing protocols such as NATS~\cite{NATS}.
External communication is handled mainly through the highly structured PostgreSQL database bindings; see Fig.~\ref{fig:FAlCon-structure}.
This separation allows computationally intensive analysis and decision-making to be performed on dedicated hardware while keeping the measurement computer lightweight and stable in the laboratory environment. In the following subsections, we describe the key abstractions of the DSL and its runtime model. 

\begin{figure}[!t]
    \centering
    \includegraphics[width=\linewidth]{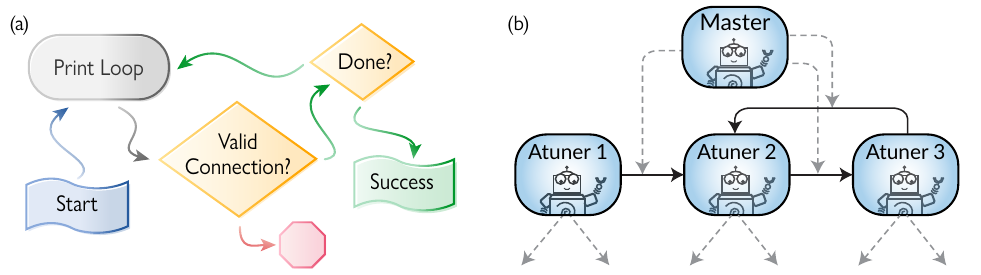}
    \caption{
    \textbf{(a)} A simplistic example of \texttt{FAlCon} DSL program visualized as a signal-flow graph.
    In this example, the program iterates through a list of connections, prints each valid connection, and terminates when the list is exhausted.
    A \texttt{FAlCon} implementation of this program is presented in Listing~\ref{lst:dsl}.
    \textbf{(b)} Illustration of hierarchical composition in \texttt{FAlCon}.
    A parent (\texttt{Master}) autotuner nests multiple child autotuners (\texttt{Atuner} 1--3) as internal states, enabling reuse by composing existing autotuners into higher-level workflows.
    }
    \label{fig:FlowAndNesting}
\end{figure}

\subsubsection{Domain-specific language}
\label{sssec:dsl}
A core design goal of \texttt{\normalfont{FAlCon}} is to express autotuning routines as explicit state-based programs that can be composed, reused, and executed reproducibly.
Many QD autotuning workflows can be represented as flowchart-like structures with conditional branching and iteration over measurement-and-analysis steps~\cite{Kovach24-BATIS}.
\texttt{FAlCon} formalizes this structure in a DSL that is designed to support hierarchical composition.
At a high level, the resulting programs naturally map to signal-flow graphs, in which nodes denote actions or subroutines and edges encode state transitions and control flow~\cite{MasonSFG}.

Figure~\ref{fig:FlowAndNesting}(a) illustrates a minimal example in which the program iterates over a list of connections and prints each valid entry.
The nodes in the flowchart represent program states\footnote{A state refers to a step in a tuning algorithm where the autotuner performs an action (e.g., measurement or analysis) or waits for information before transitioning to another state.} (e.g., \texttt{Start}, \texttt{Print Loop}) and actions (e.g., \texttt{Valid Connection?}, \texttt{Done?}) while the directed edges encode transitions and conditional branches.
Although intentionally simplistic, this minimal example reflects the same structure found in practical tuning routines.
For example, one may replace the ``Print Loop'' with a routine that acquires a charge stability diagram, evaluates a quality metric (e.g., sensor-peak visibility), and branches on whether the metric exceeds a threshold.
If the metric is below the threshold, the program updates the operating point and repeats acquisition and evaluation.
If the metric exceeds the threshold, the program transitions to a success state and advances to the next stage of the workflow.

However, the primary motivation for a dedicated DSL is not the availability of state-machine constructs, which are generally available in general-purpose languages, but the support for hierarchical composition as a first-class feature.
Figure~\ref{fig:FlowAndNesting}(b) illustrates this nesting, in which an outer autotuner (\texttt{Master}) defines a set of states whose implementations are themselves autotuners (\texttt{Atuner} 1--3) and specifies how each state transitions to the next.
This hierarchical system enables the reuse of existing tuning algorithms by simply importing them, defining them as states, and specifying how they transition to other states at a higher scope.

\subsubsection{Runtime engine}
\label{sssec:runtime-engine}
The runtime engine executes \texttt{FAlCon} DSL programs by parsing source files, constructing an internal program representation, and interpreting state transitions during execution.
The implementation, handled via the GNU Bison parser~\cite{Bison}, is designed to maintain a clear separation between parsing, program representation, and execution.
Consequently, future language updates can be implemented by modifying the parser file.

In our reference implementation, the runtime engine supports a foreign function interface (FFI), enabling routines to use external libraries exposed via C APIs.
This FFI design is particularly useful when implementing state logic in the DSL while relying on performance-sensitive analysis code or on direct measurement control libraries.
In the current architecture, interaction with the measurement system is mediated by these external interfaces rather than by embedding instrument-control code directly in the DSL.
However, it is possible to bind commonly used analysis algorithms directly via the FFI for use in future native \texttt{FAlCon} DSL code.

\subsection{\texttt{falcon-core}}
\label{sec:falcon-core}
\texttt{falcon-core} provides the foundational data types and low-level utilities used across the \texttt{FAlCon} ecosystem.
It is implemented in C++14 and is designed to support high-performance autotuning workflows for QD devices.
The repository is organized around a C++ core (\texttt{cpp/}) and a C API wrapper (\texttt{c-api/}) intended to support language bindings and external tools.
A key design goal is to provide canonical representations of common experimental concepts to enable consistent data exchange between dependent applications. 

\begin{figure}[t!]
\centering
\begin{lstlisting}[
    caption={Core definition of the \texttt{Connection} datatype in \texttt{falcon-core}. 
    A \texttt{Connection} has a name and a type (\texttt{DeviceFeature}) specifying its functional role.},
    language=C++,
    showstringspaces=false,
    abovecaptionskip=0.45\baselineskip,
    label={lst:connection-core}
    ]
namespace falcon_core::physics::device_structures {
enum class DeviceFeature {
  BarrierGate, ... 
  };
}

class Connection : public virtual generic::Song {
  std::string                     _name;
  DeviceFeature                   _type;
  mutable std::shared_timed_mutex _mu_name;
  mutable std::shared_timed_mutex _mu_type;

 public:
  // @brief Assignment operator.
  Connection& operator=(const Connection& other);
  // @brief Construct a Connection with a name and type.
  //    @param name The name of the connection.
  //    @param type The type of the connection (DeviceFeature).
  Connection(const std::string& name, const DeviceFeature& type);
  // @brief Construct a BarrierGate with a name.
  //    @param name The name of the connection.
  static std::shared_ptr<Connection> BarrierGate(const std::string& name);
  // @brief The name of the connection.
  //    @return The name as a string.
  const std::string name() const;
  // @brief The type of the connection.
  //    @return The type as a string.
  const std::string type() const;
  // @brief Check if the connection is a barrier gate.
  bool is_barrier_gate() const;  
  ...
}
\end{lstlisting}
\end{figure}

\texttt{falcon-core} includes a large collection of QD tuning-relevant data types.
An example is the \texttt{Connection}, shown in Listing~\ref{lst:connection-core}, which represents a named alias to a member of a real device's contact structure, along with its functional role (here,  barrier gate).

A key property of the \texttt{falcon-core} data types is their ability to be serialized.
All data types inherit serialization by descending from the generic \texttt{Song} class, a virtual base class that implements the \texttt{from\_json\_string} method, which can build each type directly from a JSON, and a \linebreak \texttt{to\_json\_string} method, which can turn each type directly into a string.
\texttt{Song} is implemented using the \texttt{cereal} library~\cite{Cereal}.
Composing attributes directly from Song-inheriting types ensures that the entire object is serializable.

The second key property is that these types are designed for concurrent use.
The example implementation uses shared mutexes\footnote{A mutex (mutual-exclusion lock) is a synchronization primitive that prevents race conditions by ensuring only one thread at a time can access or modify shared data.} to protect internal state during serialization; see Listing~\ref{lst:connection-serialize}.

\begin{figure}[t!]
\centering
\begin{lstlisting}[
    caption={Thread-safe serialization of \texttt{Connection}.
    The implementation uses shared mutexes to protect internal state during serialization.},
    language=C++,
    showstringspaces=false,
    abovecaptionskip=0.45\baselineskip,
    label={lst:connection-serialize}
    ]
class Connection : public virtual generic::Song {
protected:
  friend class cereal::access;
  Connection();
  template <class Archive>
  inline void serialize(Archive& ar) {
    std::shared_lock<std::shared_timed_mutex> 
        lock_name(_mu_name, std::defer_lock);
    std::shared_lock<std::shared_timed_mutex> 
        lock_type(_mu_type, std::defer_lock);
    std::lock(lock_name, lock_type);
    ar(cereal::base_class<falcon_core::generic::Song>(this), 
        _name, _type);
  }
};
\end{lstlisting}
\end{figure}

To make these data types usable from other programming environments, \texttt{falcon-core} provides a fully featured C99-compatible API alongside the C++ library.
This C API forms a hourglass pattern\footnote{A hourglass pattern is a way of writing a thin C API~\cite{ToitHIC}.}, in which C handles manage objects owned by the C++ layer.
Listing~\ref{lst:connection-c-api} demonstrates this hourglass pattern for the \texttt{Connection} datatype.
To extend the data types, developers can either add new data types to \texttt{falcon-core} or build a new library that inherits the properties of \texttt{Song}.

\begin{figure}[b!]
\centering
\begin{lstlisting}[
    caption={C-API library definition of the \texttt{Connection} datatype in \texttt{falcon-core}. 
    Each method is name-mangled by taking the C++ object type and attaching the associated method.
    The C-API maintains a handle to the shared pointer used by the underlying C++ \texttt{falcon-core} library.},
    language=C,
    abovecaptionskip=0.45\baselineskip,
    showstringspaces=false,
    label={lst:connection-c-api}
]
extern "C" {
typedef void* ConnectionHandle;
/* @category:deallocation */
void Connection_destroy(ConnectionHandle handle);
/* @category:read */
StringHandle Connection_to_json_string(ConnectionHandle handle);
/* @category:allocation */
ConnectionHandle Connection_from_json_string(StringHandle json);
/* @category:allocation */
ConnectionHandle Connection_create_barrier_gate(StringHandle name);
/* @category:read */
StringHandle Connection_name(ConnectionHandle handle);
/* @category:read */
StringHandle Connection_type(ConnectionHandle handle);
/* @category:read */
bool Connection_is_barrier_gate(ConnectionHandle handle);
...
}
\end{lstlisting}
\end{figure}

\subsection{\texttt{falcon-core-libs}}
\label{ssec:falcon-libs}
\texttt{falcon-core-libs} provides multi-language bindings to the canonical data types implemented in \texttt{falcon-core}. 
The bindings are built on the C99-compatible API exposed by \texttt{falcon-core}.
This design enables tools and services written in higher-level languages to construct, inspect, and serialize \texttt{falcon-core} data types without re-implementing their semantics or duplicating serialization logic.

The repository currently includes bindings for multiple programming environments, including Go, Python, OCaml, and Lua. 
The overall intent is to allow different parts of the \texttt{FAlCon} ecosystem to be implemented in the language that best fits their role, while preserving a shared core representation for measurement types and results.
To support evolution of the underlying C API, \texttt{falcon-core-libs} is organized so that language-specific bindings can be regenerated when \texttt{falcon-core} types and functions change\footnote{
The C API exposes functions that naturally fall into categories such as allocation, deallocation, and read/write access.
These categories provide a convenient basis for binding generators and help ensure that memory ownership and object lifetimes remain well-defined across language boundaries.}.

\subsection{\texttt{instrument-hub}}
\label{sec:instr-hub}

The \texttt{instrument-hub} serves as the central registry and discovery service within the \texttt{FAlCon} architecture, providing a unified interface for locating, querying, and addressing connected instruments; see Fig.~\ref{fig:FAlCon-structure}. 
Instantiating an instrument by the \texttt{instrument-script-server} (discussed further in \ref{sec:ins-serv}) registers it with the \texttt{instrument-hub}, exposing metadata describing the instrument, its available parameters, and the measurement quantities it can produce or consume. 

During measurement execution, the \texttt{instrument-script-server} produces measurement results as part of scheduled measurement jobs and forwards them to the \texttt{instrument-hub} in near-real time. 
The \texttt{instrument-hub} aggregates these results, manages buffering, and persists the resulting datasets to an HDF5 data store~\cite{HDF5}.

A key design principle of the \texttt{instrument-hub} is loose coupling between system components. 
Instruments are not required to know about one another at initialization time; instead, references to instruments are resolved dynamically through the \texttt{instrument-hub} when measurement routines are executed. 
This enables experimental configurations to be composed, reconfigured, and extended at runtime without restarting the control system. 
For example, a lock-in amplifier and a magnet power-supply driver may be initialized independently and later used together in a sweep measurement, with the \texttt{instrument-hub} resolving the instrument references required by the measurement workflow.

\subsection{\texttt{instrument-script-server}}
\label{sec:ins-serv}
\texttt{instrument-script-server} provides the lab-specific instrument execution and orchestration service of \texttt{FAlCon}.
It is designed to run measurement routines as scripts that reference instrument configurations, while isolating hardware-facing code from higher-level tuning logic. 
The server executes user-supplied Lua scripts, enabling experiment logic to be packaged, versioned, and replayed as structured measurement jobs.
This job-oriented execution model supports the staging and deployment of scripts, along with their required configuration and dependencies, and provides a well-defined lifecycle for running and monitoring measurements.
\texttt{instrument-script-server} manages the complete lifecycle of each instrument driver, including instantiation, initialization (e.g., opening a VISA/serial/TCP connection to the physical hardware interface), normal operation, and teardown. 
To improve robustness, the system supports process isolation by running each instrument (or driver instance) in a separate worker process, reducing the risk that a single failing driver destabilizes the entire measurement session.

The \texttt{instrument-script-server} behaves differently from other commonly used measurement systems such as \href{https://microsoft.github.io/Qcodes/}{QCoDeS} and \href{https://www.keysight.com/us/en/lib/software-detail/instrument-firmware-software/labber-3113052.html}{Labber}, which enforce a single-threaded user-facing API.
To communicate with instruments, the \texttt{instrument-script-server} runs a Lua script that performs the experiment by issuing instrument-native commands through the configured driver interfaces.

Each \texttt{instrument-script-server} instance operates as a network service that communicates with other \texttt{FAlCon} components over standard network protocols (e.g., TCP/IP). 
This enables deployments in which the server runs on a dedicated laboratory computer physically connected to the instruments, while the experimentalist interacts with the system from a separate workstation or remote location.

Beyond single-instrument operations, the \texttt{instrument-script-server} provides built-in support for coordinated multi-instrument measurements.
The server supports parallel command execution and synchronization primitives, enabling scripts to align actions across instruments and maintain consistent timing relationships.
Results and execution metadata are returned to upstream consumers through the server's integration interfaces as part of a traceable measurement record.

A common measurement pattern is a parameter sweep, in which one or more independent variables are stepped through specified ranges while one or more dependent variables are recorded at each point.
Within \texttt{instrument-script-server}, such measurement patterns are implemented as scripts and benefit from centralized orchestration, parallelism, and synchronization across participating instruments.
Because the execution layer is centralized, scripts do not need to re-implement low-level driver management or ad hoc multi-instrument coordination logic.

\subsection{\texttt{falcon-measurement-lib}}
\label{sec:measurement-lib}
\texttt{falcon-measurement-lib} provides the schema-driven measurement interface layer of \texttt{FAlCon}.
Its central goal is to standardize how measurement requests, script contexts, and returned data products are represented across heterogeneous laboratory setups.
It defines canonical JSON schemas\footnote{JSON Schema is an open, IETF-recognized standard language that declaratively defines the structure and content of JSON data for validation, documentation, and interaction control.} for reusable types (e.g., instrument targets and domains) as well as per-script schemas that specify the expected global variables and context available to a measurement script.
This schema-first approach enables validation of measurement requests and improves portability by making measurement interfaces explicit rather than implicit in ad hoc scripts.

In addition to the schema definitions, \texttt{falcon-measurement-lib} provides code-generation tools that automatically produce helper files such as editor type hints and language-specific type definitions.
These generated outputs include Emmy-style typing information for editor integration, Go struct types for compiler and tooling support, and optional Teal scaffolds to enable typed scripting workflows.
\texttt{falcon-measurement-lib} also provides Lua runtime helper modules that implement schema-referenced functionality and support script execution.
Together, these components maintain consistent measurement interfaces across the \texttt{FAlCon} ecosystem and reduce integration errors when extending the measurement library or adding new scripts.

\texttt{falcon-measurement-lib} defines measurement templates to facilitate portable data acquisition across the \texttt{FAlCon} ecosystem.
Each schema in the library defines templated measurement styles.
An example \texttt{set-get} measurement is shown in Listing~\ref{lst:get-set-measurement-schema}.
This example captures a common pattern in which one or more instrument parameters are set, and one or more values are then acquired and returned.
Such templates provide a portable vocabulary for expressing measurement intent that can be compiled or executed on lab-specific instrument configurations through the rest of the \texttt{FAlCon} stack.

\begin{figure}[t!]
\centering
\begin{lstlisting}[
    caption={A code snippet for scripts that perform DC get/set operations. 
    It allows setting voltages on specified instruments and reading values from others, supporting batch operations, parallel execution, and logging. This context is typically used for scripts that need to set outputs and acquire data points from instruments in a coordinated way.},
    language=json,
    abovecaptionskip=0.45\baselineskip,
    showstringspaces=false,
    label={lst:get-set-measurement-schema}
]
{
  "title": "Measure_Get_Set",
  "properties": {
    "setVoltages": {
      "additionalProperties": {
        "type": "number"
      },
      "description": "Table mapping instrument target identifiers (as strings) to voltage values to set in V.",
      "type": "object"
    },
    "sampleRate": {
      "description": "Sampling rate in Hz. Defines the rate at which waveform data is sampled during acquisition.",
      "type": "number"
    },
    "numPoints": {
      "descriptions": "Number of points to acquire per datapoint. Specifies the length of each acquired waveform.",
      "type": "integer"
    },
    "getters": {
      "description": "Instruments to read from.",
      "type": "array"
    },
    "setters": {
      "description": "Instruments to write to.",
      "type": "array"
    }
  },
  "returns": {
    "type": "array",
    "description": "List of MeasurementResponse with a numeric value in nA.",
    }
  }
}
\end{lstlisting}
\end{figure}

\section{Designing Algorithms} 
\label{sec:FAlCon-alg}
Algorithms are expressed in the \texttt{FAlCon} DSL.
The DSL represents tuning routines as state-based programs with named states, typed inputs and outputs, and well-defined transition logic.
This design aligns with the dominant structure of many QD tuning workflows, which naturally decompose into staged procedures with conditional branching, retries, and termination criteria.
The DSL isolates tuning intent from lab-specific software details, reducing incidental complexity and making routines easier to exchange across groups.
It also provides a controlled surface for extension, allowing \texttt{FAlCon} to evolve language features and execution backends as requirements change.

\subsection{Data structure definitions}
\label{sec:alq-dsd}

\begin{figure}[t!]
\centering
\begin{lstlisting}[
    caption={Example illustrating core \texttt{FAlCon} DSL constructs, including imports, user-defined data types, and an \texttt{autotuner} expressed as a state-based program with explicit transitions and error handling.},
    language=falconCode,
    abovecaptionskip=0.45\baselineskip,
    showstringspaces=false,
    label={lst:dsl}
]
import ("log" "io" "config" "array")
struct DeviceConnection {
    string name_;
    routine New(string name) -> (DeviceConnection conn) {
        name_ = name;
    }
    routine Name() -> (string name) {
        name = this.name_;
    }
}
autotuner CheckPlungerGates (array::Array<DeviceConnection> connections,
                             config::Config conf) -> (Error err) {
    int index = 0;
    err = nil;
    DeviceConnection connection = connections[index];
    start -> loop (connection);
    state loop (DeviceConnection conn) {
        io::println("DeviceConnection name: " + conn.Name());
        if (!conf.GetPlungerGates().Contains(conn)) {
            log::warn("DeviceConnection " + conn.Name() +
                      " is not in plunger gates.");
            -> missing_plunger_gate;
        } elif (index < connections.Size()) {
            index = index + 1;
            -> loop(connections[index]);
        } else {
            -> exit;
        }
    }
    state missing_plunger_gate (DeviceConnection conn) {
        err = Error("DeviceConnection " + conn.Name() +
                    " is missing a plunger gate.");
        -> exit;
    }
    state exit {
        log::info("Finished processing connections.");
        terminal;
    }
}
\end{lstlisting}
\end{figure}

Listing~\ref{lst:dsl} provides a compact, primarily pedagogical example that illustrates the \texttt{FAlCon} DSL constructs and the state-machine programming model used throughout \texttt{FAlCon}.
The program begins with an \texttt{import} block that loads required libraries (\codelines{1}).
Only libraries native to \texttt{FAlCon} are supported. 
In \codelines{2}{10}, a new data structure \texttt{DeviceConnection} is defined.
\texttt{DeviceConnection} contains a single attribute called \texttt{name\_} and two associated routines: a method called \texttt{New} and a routine to access the attribute \texttt{Name}.
A custom \texttt{autotuner}, \texttt{CheckPlungerGates}, implementing a simple check over a list of device connections and demonstrating conditional branching and error handling within the state-machine model, is defined in \codelines{11}{39}.
\texttt{CheckPlungerGates} is a demonstration of a finite-state program with typed inputs and outputs, local state variables (\texttt{index}, \texttt{connection}, and \texttt{err}), and an explicit start transition into the main \texttt{loop} state.
Within \texttt{loop}, the autotuner iterates over device connections, prints each entry, and branches on conditional logic, either transitioning to \texttt{missing\_plunger\_gate} to construct and propagate an error or advancing the loop to process the next connection (\codelines{17}{29}).
The routine terminates in an explicit \texttt{exit} state marked \texttt{terminal}.

\begin{figure}[b!]
\centering
\begin{lstlisting}[
    caption={
        An example autotuner that can change the charge state of a quantum dot device.
        This example is simple since it is not defined from a general device configuration file. 
        The autotuner uses the \texttt{BlipStateStepper} routine, which is guaranteed to step a single charge state in a specified direction.
        },
    language=falconcode,
    abovecaptionskip=0.45\baselineskip,
    showstringspaces=false,
    label={lst:stint-code}
]
import ("stateStepper" "deviceStates" "hub")
// (n,m) indicates the x,y axis count
autotuner ChargeConfigurationTuner (int n, int m) -> (deviceStates::DeviceStates dstates) {
    // this autotuner starts the device tuned at (0,0)
    int current_n=0;
    int current_m=0;
    // this direction can either be up or right
    string startingDirection = "up";
    start -> tuning (startingDirection);
    state tuning (string direction) {
        stateStepper::BlipStateStepper(direction);
        if (direction == "up") {
            current_n = current_n + 1;
            if (current_n < n) {
                -> tuning("up");
            } else {
                -> tuning("right");
            }
        } else {
            current_m = current_m + 1;
            if (current_m < m) {
                -> tuning("right");
            } else {
                dstates = hub::CollectCurrentDeviceState();
                terminal;
            }
        }
    } 
}
\end{lstlisting}
\end{figure}

\subsection{Charge configuration navigation for a double quantum dot system}
\label{sec:alq-cct}
A more practical example is presented in Listing~\ref{lst:stint-code}, which gives a simple program for performing charge-configuration navigation.
The \texttt{ChargeConfigurationTuner} defines a simple yet correct autotuner for navigating to the $(n,m)$ configuration in charge-stability space, implicitly assuming navigation of a double-QD system in two Cartesian dimensions corresponding to the sweep gates.
Algorithmically, the \texttt{ChargeCongfigurationTuner} is equivalent to a simple counter, and uses a direction variable and current $n$ and $m$ variables to navigate to an implicit \texttt{terminal} state.

The program begins with an \texttt{import} block that loads required \texttt{stateStepper}, \texttt{deviceState}, and \texttt{hub} libraries.
The \texttt{stateStepper} library implements a \texttt{BlipStateStepper} autotuner that sweeps the voltages of the plunger gates of the QD device in specified \texttt{direction} to detect the first peak in the conductance (a \textit{blip}).
It then places the QD device in the center of the charge configuration cell past that blip.
The \texttt{right}(\texttt{up}) direction indicates blip detection along a gate-voltage sweep in the rightward(upward) direction, commonly associated with the faster (slower) sweep axis of the charge stability diagram.
The \texttt{deviceStates} library implements the \texttt{DeviceStates} data type that describes the current voltage configuration of an entire QD device.
This is broken down into constituent parts consisting of a list of \texttt{DeviceState}, where each \texttt{DeviceState} contains a reference to the associated \texttt{Connection}\footnote{See Listing~\ref{lst:connection-core} for definition.} (in this case, a \texttt{PlungerGate}) and a \texttt{Quantity} representing the voltage applied to this gate with a float and a \texttt{SymbolUnit} indicating measurement units (here, \si{volt}).
Finally, the \texttt{hub} library implements a binding to the \texttt{instrument-script-server} and allows the autotuner to request the current \texttt{DeviceState} directly from the instruments using the routine \texttt{CollectCurrentDeviceState}.

The \texttt{ChargeConfigurationTuner} takes as input two integers, $m$ and $n$, that describe the user-requested target charge configuration of the system (\codelines{3}).
The autotuner terminates when it reaches its target charge configuration, returning the current \texttt{deviceStates} of the system.
Initially, the \texttt{autotuner} assumes that the device is tuned into the $(0,0)$ charge configuration.
This is encoded as two local \texttt{autotuner} state variables, called \texttt{current\_m} and \texttt{current\_n} (\codelines{5}{6}).
These variables count the blips along the two cardinal measurement directions, with the default starting direction set to \texttt{up} (\codelines{8}).

Once the initial configuration is complete, the \texttt{ChargeConfigurationTuner} state changes from \texttt{start} to \texttt{tuning}, with the initial measurement direction passed into the local state variables defined inside of the \texttt{tuning} state and assigned to the variable name \texttt{direction}.
Once inside \texttt{tuning}, the \texttt{ChargeConfigurationTuner} calls a \texttt{stateStepper::BlipStateStepper} subroutine to drive the QD device into the consecutive cell in the direction passed into the subroutine (\codelines{11}). 

The \texttt{autotuner} state variable \texttt{current\_n} is then incremented and checked against the target charge configuration for this direction (\codelines{13}{14}). 
If the target is not met, the current \texttt{tuning} state transitions to a new \texttt{tuning} state in the \texttt{up} direction (\codelines{15}).  
Otherwise, the current \texttt{tuning} state transitions to a new \texttt{tuning} state in the \texttt{right} direction (\codelines{16}{18}).
The measurements in the \texttt{right} direction continue until the \texttt{autotuner} state variable \texttt{current\_m} agrees with the target $m$ \texttt{autotuner} state variable. 

At that point, the \texttt{ChargeConfigurationTuner} proceeds to call from the \texttt{hub} library the \texttt{CollectCurrentDeviceState} subroutine, which builds the updated \texttt{DeviceStates} (\codelines{24}).
The \texttt{DeviceStates} is then stored into the \texttt{autotuner} state variable called \texttt{dstates} that is expected on the output after successful completion of the \texttt{autotuner} execution (\codelines{25}).

\section{Conclusion}
\label{sec:conclusion}
As QD arrays grow in scale and complexity, reproducible, automated, and portable control workflows become increasingly important.
Efficient characterization and operation of QD systems at scale will soon require universal, automated control routines that can be shared, adapted, and reused.
\texttt{FAlCon} lays the groundwork for such workflows by standardizing data representations, enabling typed, validated scripting, and offering an extensible architecture amenable to expansions of device layouts, measurement schemas, and instrument backends.

\texttt{FAlCon} is an open-source software ecosystem designed to make characterization and autotuning routines easier to share and deploy across heterogeneous device platforms, control electronics, and qubit types.
It addresses the portability barrier by separating algorithm intent from instrument realization, combining a state-based DSL for tuning logic with standardized, physics-informed data structures and a schema-driven measurement interface.
Together, these components enable validated, typed scripting and a consistent way to represent measurement requests, contexts, and data products across devices and laboratories.

The \texttt{FAlCon} ecosystem achieves expressive state-based tuning logic via its DSL.
Separating state-based tuning logic from the measurement control backend makes this approach hardware-agnostic.
This separation is enabled by a common library of communicable, physics-informed QD data types.
The measurement stack uses libraries of shared measurement protocols to robustly control full experimental setups.
Taken together, these design choices allow \texttt{FAlCon} to interface with instruments across a wide range of laboratory configurations.
As a developing full-stack ecosystem, \texttt{FAlCon} aims to bridge the gap between system engineering and experimental deployment in state-of-the-art quantum dot research.

Looking ahead, we expect community contributions to expand the ecosystem along several axes.
Near-term extensions include additional vernacular types and measurement templates, broader instrument support, and libraries of reusable autotuners that can be benchmarked across devices.
Constructing and sharing explicit benchmarking simulations will enable direct comparisons of algorithms' performance and efficiency, regardless of instrumentation-induced bottlenecks, providing end users with the most performant algorithms available.
More ambitious directions include integrating simulation and digital twin workflows for offline debugging and evaluation of tuning strategies, adding machine learning decision layers as optional components within tuning routines, and exploring compiling \texttt{FAlCon} measurement commands directly to instrument hardware ISAs---bypassing the current manual Lua scripting layer in the \texttt{instrument-script-server}.
As control electronics move away from distributed, standalone laboratory instruments toward scalable on-chip solutions, work could also be done to compile the \texttt{FAlCon} engine directly for a field-programmable gate array (FPGA).

As an open-source software project, \texttt{FAlCon} establishes an active interdisciplinary link between quantum physics research and software algorithm development.
By providing a shared software foundation for algorithmic control, \texttt{FAlCon} aims to catalyze faster iteration, improved reproducibility, and more systematic comparison of tuning strategies across laboratories.
We anticipate that such capabilities will help reduce duplicated engineering effort and support the transition toward larger, more autonomous QD-based quantum processors.

Beyond immediate applications in semiconductor spin qubits, the abstractions introduced in \texttt{FAlCon} are designed to generalize to other scientific platforms, supporting a broader vision of interoperable autotuning across fields of need.
The ecosystem is designed to support users across a range of expertise, from those running pre-built algorithms with custom initial conditions, to contributors developing new autotuning routines, to developers extending instrumentation support and upgrading the platform itself.
The only requirement is that users simplify their problem into nested state machines.

\section*{Acknowledgements}

\paragraph{Funding information}
This work was supported in part by ARO Grants No. W911NF-17-1-0274, No. W911NF-23-1-0110, and No. W911NF-24-2-0043.
This research was performed in part while Z.M. held an NRC Research Associateship award at NIST.
The views and conclusions contained in this paper are those of the authors and should not be interpreted as representing the official policies, either expressed or implied, of the U.S. Government. 
The U.S. Government is authorized to reproduce and distribute reprints for Government purposes, notwithstanding any copyright noted herein. 
Any mention of commercial products is for information only; it does not imply recommendation or endorsement by NIST.

\paragraph{Software availability} 
\texttt{FAlCon}'s binaries and installation documentation are accessible via the \href{https://github.com/falcon-autotuning}{FAlCon} GitHub organization.
If you find this software useful, please star a specific repository and cite this paper.



\end{document}